\documentclass[
  journal=apchd5,
  manuscript=article,
  layout=traditional,
  email=true,
  keywords=true,
  super=true
]{achemso}

\usepackage[version=4]{mhchem}
\usepackage{graphicx}
\usepackage{siunitx}
\usepackage{overpic}   
\usepackage{float}     
\usepackage{cleveref}
\usepackage[dvipsnames]{xcolor}
\usepackage[T1]{fontenc}

\author{Maninder Kaur}
\affiliation[University of Georgia]
{Department of Physics and Astronomy, University of Georgia, Athens, GA, USA}
\author{Nicolas T. Sandino}
\affiliation[University of Georgia]
{Department of Physics and Astronomy, University of Georgia, Athens, GA, USA}
\author{Jason P. Terry}
\affiliation[University of Georgia]
{Department of Physics and Astronomy, University of Georgia, Athens, GA, USA}
\author{Mahdi Ghafariasl}
\affiliation[University of Georgia]
{Department of Physics and Astronomy, University of Georgia, Athens, GA, USA}
\author{Yohannes Abate}
\email{yohannes.abate@uga.edu}
\affiliation[University of Georgia]
{Department of Physics and Astronomy, University of Georgia, Athens, GA, USA}

\title
  {Excitonic Landscapes in Monolayer Lateral Heterostructures Revealed by Unsupervised Machine Learning}


\begin{document}

\begin{abstract}

Two-dimensional (2D) in-plane heterostructures including compositionally graded alloys and lateral heterostructures with defined interfaces display rich optoelectronic properties and offer versatile platforms to explore one-dimensional interface physics and many-body interaction effects. 
Graded \(\mathrm{Mo}_x\mathrm{W}_{1-x}\mathrm{S}_2\) alloys show smooth spatial variations in composition and strain that continuously tune excitonic emission, while \(\mathrm{MoS}_2\)--\(\mathrm{WS}_2\) lateral heterostructures contain atomically sharp interfaces supporting one-dimensional excitonic phenomena. These single-layer systems combine tunable optical and electronic properties with potential for stable, high-performance optoelectronic devices. Hyperspectral and nano-resolved photoluminescence (PL) imaging enable spatial mapping of optical features along with local variations in composition, strain, and defects, but manual interpretation of such large datasets is slow and subjective. 
Here, we introduce a fast and scalable unsupervised machine-learning (ML) framework to extract quantitative and interpretable information from hyperspectral PL datasets of graded \(\mathrm{Mo}_x\mathrm{W}_{1-x}\mathrm{S}_2\) alloys and \(\mathrm{MoS}_2\)--\(\mathrm{WS}_2\) heterostructures. Combining principal-component analysis (PCA), t-distributed stochastic neighbor embedding (t-SNE), and density-based spatial clustering (DBSCAN), we uncover spectrally distinct domains associated with composition, strain, and defect variations. Decomposition of representative spectra reveals multiple emission species, including band-edge excitons and defect-related transitions, demonstrating that ML-driven analysis provides a robust and automated route to interpret rich optical properties of 2D materials.
\end{abstract}


\section{Introduction}
Two-dimensional (2D) materials have remarkable electronic, optical, and mechanical properties, including layer-dependent bandgaps, exciton physics, strain-induced band modifications, and defect-bound states. Among these, transition metal dichalcogenides (TMDCs) such as MoS$_2$ and WS$_2$ form a particularly important class, widely investigated for their excitonic and defect-related properties~\cite{mak2010atomically, Hong2015, Mak2016, wang2018colloquium,he2013experimental,tongay2013defects,Yin2021_CSR}. 

Compositionally graded 2D alloys such as Mo$_x$W$_{1-x}$S$_2$ are of great importance because they enable precise control of electronic and optical properties within a single crystal~\cite{ghafariasl2024}. By gradually varying the ratio of Mo and W across the layer, the alloy allows continuous tuning of the bandgap and excitonic transitions, bridging the properties of pure MoS$_2$ and WS$_2$. This built-in compositional gradient also produces smooth variations in strain and defect concentration, giving rise to localized changes in exciton energy and intensity. Such spatial control over the band structure and defect landscape provides an ideal platform for exploring how composition, strain, and vacancies together shape light–matter interactions, exciton dynamics, and charge transfer in 2D semiconductors~\cite{ghafariasl2024,he2013experimental,tongay2013defects}.

In-plane (lateral) TMDC heterostructures represent another key platform for exploring nanoscale optical and electronic behavior. Unlike graded alloys, these systems contain atomically sharp covalent interfaces between dissimilar monolayer domains that form through seamless lattice stitching during growth~\cite{li2015epitaxial,fali2021photodegradation}. Such one-dimensional junctions give rise to unique interface physics, including exciton diffusion, charge-carrier funneling, and strain relaxation across the boundary~\cite{huang2014lateral,fali2021photodegradation}. The presence of alloyed transition regions at the interface can also stabilize the structure against oxidation and photodegradation, providing long-term optical and chemical stability under illumination~\cite{komsa2012two,fali2021photodegradation}. These lateral heterostructures thus serve as ideal building blocks for planar monolayer \emph{p--n} junctions, excitonic circuits, and high-mobility field-effect transistors, while offering a platform to investigate one-dimensional confinement, defect-assisted recombination, and interface-driven exciton dynamics in 2D semiconductors~\cite{najmaei2013vapour}. Advanced imaging and spectroscopy techniques, including PL mapping, scanning near-field optical microscopy (SNOM), and other scanning probe methods, enable spatially resolved visualization of such phenomena at nanometer scales~\cite{lee2012synthesis,fei2012gate,najmaei2013vapour,Abate2016Nanoscopy,fali2021photodegradation,Hillenbrand2025,GarciadeAbajo2025Roadmap}. However, these experiments generate enormous hyperspectral datasets, making manual analysis and interpretation challenging. 

Machine learning (ML) has become a crucial part of modern scientific data analysis, providing tools for dimensionality reduction, clustering, and pattern recognition in complex datasets~\cite{Ryu2022ML2Dmaterials,Butler2018,vonLilienfeld2020,Schmidt2019,Zhu2024_AutonomousSTM,Chen2021_HML_sSNOM,Imamura2024,Fu2024_sparseSNOM,Khor2023,jia2022ml_spectral_nanomat,lu2024when,ElAbbassi2021_Unsupervised,Kolesnichenko2021_ExcitonicSpectra, He2025_OpticalID2D,roter2022clustering}. ML techniques have been successfully used to identify phase transitions~\cite{Carrasquilla2017,Wetzel2017}, discover order parameters~\cite{Wang2016}, and analyze diverse spectroscopic and imaging data, including scanning tunneling microscopy, near-field optical microscopy, and PL studies of 2D
materials~\cite{Zhu2024_AutonomousSTM,Chen2021_HML_sSNOM,Imamura2024,Fu2024_sparseSNOM,Khor2023,Djordevic2018,Tanaka2019,Ramprasad2017,jia2022ml_spectral_nanomat,ElAbbassi2021_Unsupervised,Kolesnichenko2021_ExcitonicSpectra,Ryu2022ML2Dmaterials,He2025_OpticalID2D,roter2022clustering,Mao2020_RamanML}. Recent developments demonstrate that ML can classify multiple 2D materials from Raman or PL spectra with high accuracy, underscoring its ability to handle complex experimental data~\cite{qi2024deep,Lu2022_Unraveling,lu2024when,Joung2021_JACSAu_DLOptSpec,Choo2018,Kalantre2019,Carleo2019_RevModPhys, Frey2020}. 

In this work, we apply PCA, t-SNE, DBSCAN clustering, and multi-peak Gaussian fitting to hyperspectral PL datasets of graded Mo$_x$W$_{1-x}$S$_2$ monolayers and MoS$_2$--WS$_2$
lateral heterostructures. Our analysis reveals spatially distinct optical domains corresponding to composition gradients, strain-induced spectral shifts, and defect-related emission. By quantitatively decomposing individual spectra, we establish a direct connection between unsupervised ML clusters and physical material properties, offering a comprehensive strategy to characterize nanoscale optical heterogeneity in 2D materials.

\section*{Methods}
Graded monolayer $\mathrm{Mo}_x\mathrm{W}_{1-x}\mathrm{S}_2$ alloys were synthesized using an alkali-metal-halide-assisted chemical vapor deposition (CVD) process that produces a continuous lateral gradient in composition from $\mathrm{Mo}$-rich centers to $\mathrm{W}$-rich edges~\cite{ghafariasl2024}. 
The growth utilized $\mathrm{MoO_3}$ and $\mathrm{WO_3}$ powders as precursors, elemental sulfur as the chalcogen source, and $\mathrm{NaCl}$ as a growth promoter. The reaction was carried out at 750–800~$^\circ$C on $\mathrm{SiO_2/Si}$, yielding large-area monolayers with smooth compositional transitions.
In-plane $\mathrm{MoS_2}$–$\mathrm{WS_2}$ heterostructures were grown by a sequential CVD process in which the Mo and W oxide precursors were introduced at different stages of the reaction to promote lateral epitaxy~\cite{fali2021photodegradation}. Initially, $\mathrm{MoS_2}$ domains nucleated on the $\mathrm{SiO_2/Si}$ substrate at $\sim$750~$^\circ$C under sulfur vapor. Once the $\mathrm{MoS_2}$ growth was complete, the Mo precursor was removed, and $\mathrm{WO_3}$ vapor was introduced to drive edge-selective substitutional growth of $\mathrm{WS_2}$ at the periphery of existing $\mathrm{MoS_2}$ crystals. This process yielded laterally stitched monolayer heterostructures with atomically sharp junctions and narrow alloy transition regions.
Hyperspectral PL data cubes were then acquired using a scattering-type near-field optical microscope (s-SNOM) integrated with a broadband excitation source. 
Each hyperspectral cube contained $10^3$ spectral channels per pixel, with a nanoscale spatial resolution. The $x$ and $y$ axes of the cube correspond to the lateral dimensions of the sample surface, while the $z$-axis represents the photon energy range ($1.85$–$2.0~\text{eV}$). Each spectrum was acquired with a $1~\text{s}$ integration time, resulting in a total acquisition time of approximately $2~\text{h}$ per image. No additional denoising or normalization was applied beyond experimental preprocessing; all analyses were performed directly on the acquired hyperspectral cube $I(x, y, E)$.

\par
\begin{figure}[H]
    \centering
    \includegraphics[width=\linewidth]{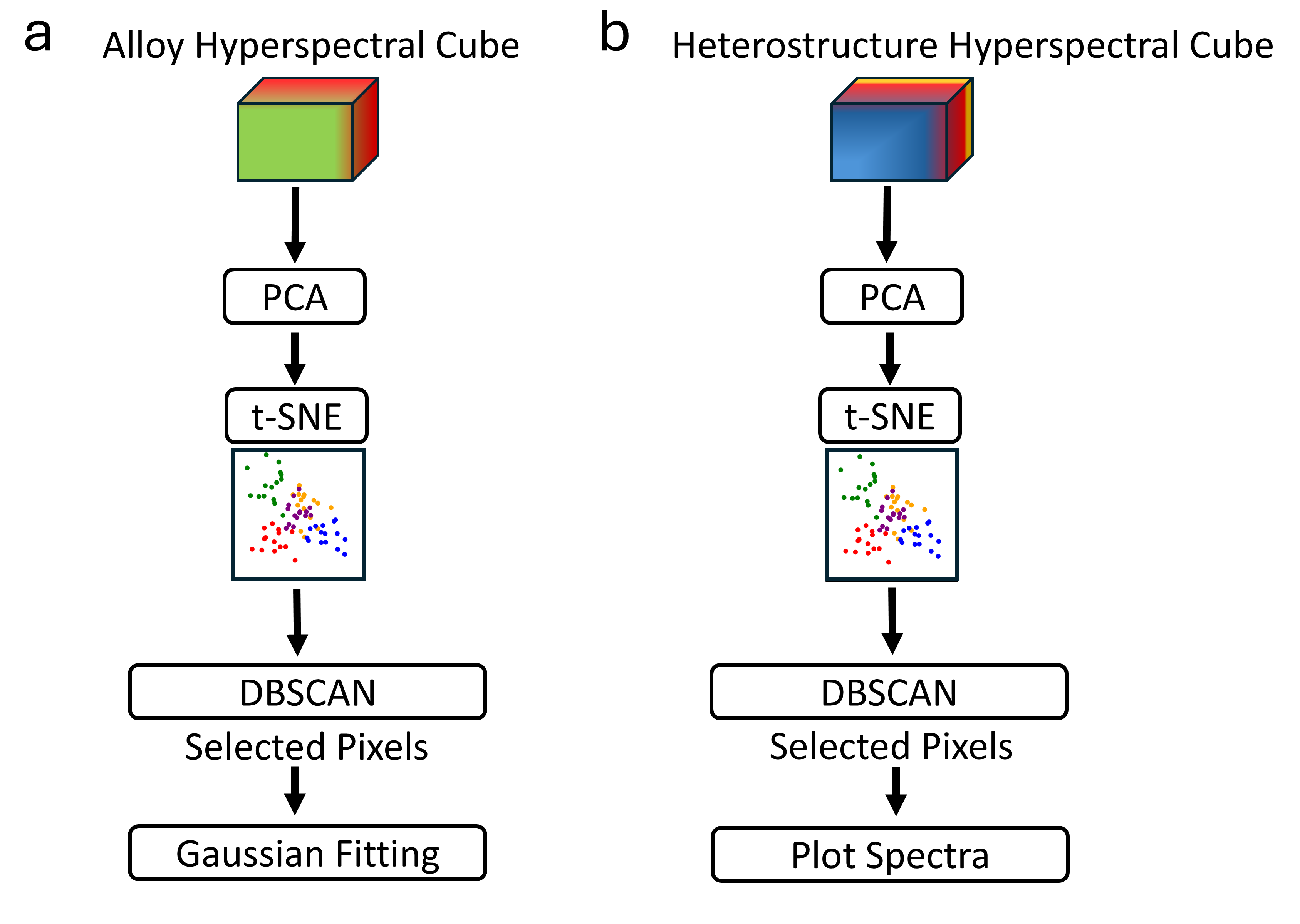}
\caption{\textbf{Workflow for unsupervised ML analysis of hyperspectral PL data from alloy and heterostructure monolayers.}
(a) For alloy samples, the hyperspectral cube is reduced using PCA, followed by t-SNE visualization and DBSCAN clustering; selected pixels are then subjected to Gaussian fitting.
(b) For heterostructure samples, the same dimensionality-reduction and clustering workflow is applied.}
\label{fig:method}
\end{figure}

For ML analysis, each hyperspectral cube was reshaped so that individual pixel spectra formed rows in a
two-dimensional array of size $(H \cdot W) \times C$, 
where $H$ and $W$ are the spatial dimensions and $C$ is the number of spectral channels. 
For the alloy dataset ($75 \times 75 \times 1024$), this produced a matrix of $75^2 \times 1024 = 5625 \times 1024$. For the heterostructure dataset 
($48 \times 48 \times 1024$), the resulting matrix was $48^2 \times 1024 = 2304 \times 1024$. This representation enabled PCA to reduce the high-dimensional spectral data and identify the main modes of variation, i.e., principal components (PCs).
The PCA-reduced data were then visualized using t-SNE to project the spectral relationships into two dimensions, allowing intuitive assessment of similarities and differences across the sample. For unsupervised clustering, both the alloy and heterostructure datasets were analyzed using 
DBSCAN applied to the PCA-reduced spectra. The complete workflow of the ML analysis is illustrated in~\Cref{fig:method}. In each case, the data were first reduced to three principal components and then embedded into a two-dimensional t-SNE space 
using a fixed random seed for reproducibility. DBSCAN clustering was 
performed with $\mathit{eps} = 2.5$ and $\mathit{min\_samples} = 6$ to group pixels according to 
their spectral similarity in the reduced feature space. For the alloy dataset, representative spectra 
from each cluster were modeled using one-, two-, or three-component Gaussian fits to decompose the 
PL emission into individual peaks. Parameters such as the amplitude and center wavelength of the dominant emissive state were extracted from the best-fitting models and spatially mapped to establish correlations between the emission properties and the machine-learning–identified clusters.
For the heterostructure dataset, representative pixels from each cluster were directly plotted to visualize raw spectral differences across the dataset.

\section{Results and discussion}
\subsection{Unsupervised  ML analysis of graded alloyed 2D $\mathrm{Mo}_x\mathrm{W}_{1-x}\mathrm{S}_2$}
\begin{figure}[tb]
    \centering
    \includegraphics[width=\linewidth]{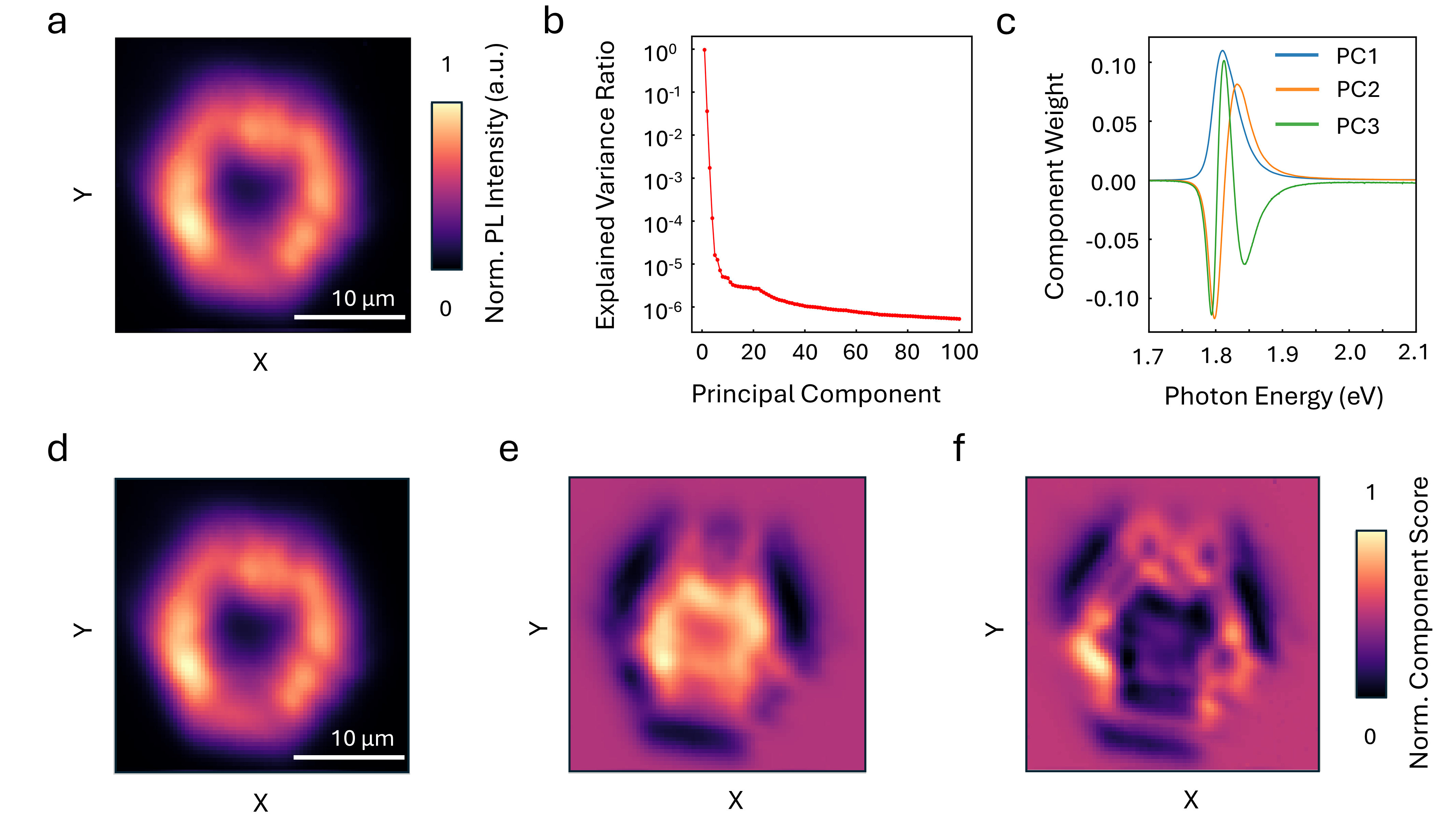}
\caption{\textbf{Hyperspectral analysis of a $\mathrm{Mo}_x\mathrm{W}_{1-x}\mathrm{S}_2$ monolayer.}
(a) Total PL intensity map obtained directly from experimental data (without machine learning). 
(b) PCA explained variance plot showing the fraction of total variance captured by each component. 
(c) Eigen-spectra ($\mathit{PC1}$–$\mathit{PC3}$) corresponding to the main modes of spectral variation. 
(d–f) Spatial projections of ($\mathit{PC1}$–$\mathit{PC3}$) across the flake.}
\label{fig:pca_image_alloy}
\end{figure}

The analysis in this section is based on graded $\mathrm{Mo}_x\mathrm{W}_{1-x}\mathrm{S}_2$ alloyed monolayer experimental data acquired using tip-based PL mapping in an s-SNOM.~\cite{ghafariasl2024}
\Cref{fig:pca_image_alloy} provides a full overview of how the PL emission varies across a graded $\mathrm{Mo}_x\mathrm{W}_{1-x}\mathrm{S}_2$ monolayer, both in terms of total intensity and in the underlying spectral features extracted using PCA. It combines a raw intensity map, a summary of spectral variation, and spatial distribution of the top three principal components. \Cref{fig:pca_image_alloy}(a)  shows the total experimental PL intensity map which is acquired by integrating each pixel’s spectrum over all photon
energies, and is used as a reference for evaluating spatial patterns in PCA-derived components.  This map reveals that the monolayer has a compositional gradient, with the Mo-rich region near the center exhibiting lower PL intensity and the alloy/W-rich transition zones at the periphery showing stronger emission, resulting in a tunable bandgap of $1.85$ to $1.95~\text{eV}$ when moving from the center to the edge of the flake~\cite{ghafariasl2024}. We then perform PCA to reduce the dimensionality of the spectral dataset and extract the main types of variation across the sample, shown in \Cref{fig:pca_image_alloy}(b–f). This analysis helps us break down the complex dataset into just a few patterns that are easier to visualize and understand. \Cref{fig:pca_image_alloy}(b) shows how much variance each PC captures. The first three PCs together account for more than $99\%$ of the total variance, indicating that they capture nearly all meaningful spectral variations in the dataset. This allows the analysis to be restricted to $\mathit{PC1}$, $\mathit{PC2}$, and $\mathit{PC3}$ for identifying major trends in the PL spectra, rather than working with hundreds of individual spectral channels.

\Cref{fig:pca_image_alloy}(c) shows the spectral weight vectors (also called eigen-spectra) for $\mathit{PC1}$, $\mathit{PC2}$, and $\mathit{PC3}$. These describe what each PC is looking for in the spectra. $\mathit{PC1}$ has a strong positive peak near $1.82~\text{eV}$ and resembles the average PL spectrum. $\mathit{PC2}$ changes sign around $1.82~\text{eV}$, meaning it reacts to whether the peak shifts to lower or higher energy. $\mathit{PC3}$ has a dip near the higher-energy shoulder. These spectral shapes give physical meaning to the PCs: $\mathit{PC1}$ relates to total intensity, $\mathit{PC2}$ detects shifts in the emission peak (due to alloying, defects, or strain), and $\mathit{PC3}$ highlights more subtle differences in shape, possibly from defects or strain. \Cref{fig:pca_image_alloy}(d) shows where $\mathit{PC1}$ is strong or weak across the sample. It looks nearly identical to the total intensity map in Figure~2(a). This similarity confirms that $\mathit{PC1}$ mainly tracks overall PL intensity, so it can be used as a simpler version of the full intensity map. \Cref{fig:pca_image_alloy}(e) shows the spatial pattern of $\mathit{PC2}$. There is a clear contrast between center and edges.
This contrast matches the spectral weight vector of $\mathit{PC2}$, which flips sign near the peak energy. This means that in the center of the sample, the PL peak is slightly red-shifted (lower energy), while toward the edge it is blue-shifted (higher energy). $\mathit{PC2}$ therefore visualizes how the exciton emission spectra change from $\mathrm{Mo}$-rich to $\mathrm{W}$-rich regions. \Cref{fig:pca_image_alloy}(f) shows where $\mathit{PC3}$ is active. Unlike the other two, this map reveals finer features---small patches that might be linked to defects or local strain. The detailed patterns in $\mathit{PC3}$ suggest it detects subtle changes in spectral shape, which could come from non-uniform strain or defect distributions. While the spectral weight vectors tell us what kind of spectral feature each PC responds to, the spatial maps tell us where those features appear on the sample. Both are important because they connect physical causes to their spatial effects.

\begin{figure}[tb]
    \centering
    \includegraphics[width=\linewidth]{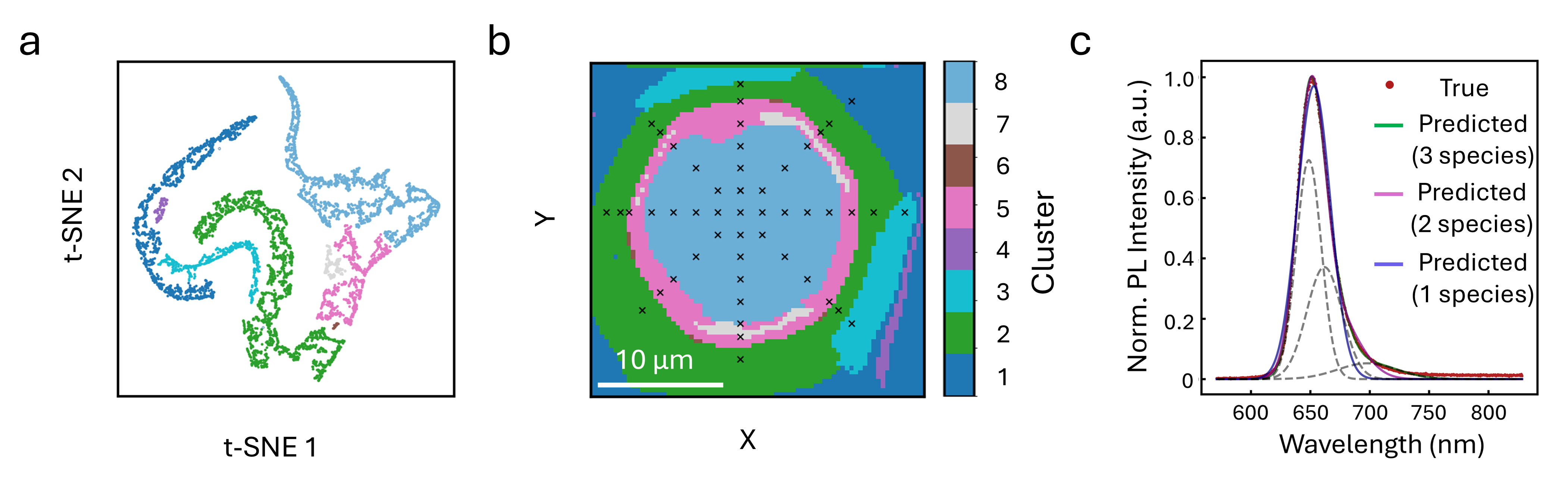}
\caption{\textbf{Unsupervised clustering and visualization of hyperspectral PL spectra from a graded $\mathrm{Mo}_x\mathrm{W}_{1-x}\mathrm{S}_2$ monolayer.}
(a) t-SNE projection of the PCA-reduced spectra colored by DBSCAN cluster labels. 
(b) Real-space mapping of the same clusters across the flake. 
(c) Representative Gaussian fits to a selected pixel spectrum (from the 51 marked in panel~b), showing one-, two-, and three-component fitting models for comparison.}
\label{fig:alloy_dbscan_results}
\end{figure}

\Cref{fig:alloy_dbscan_results} aims to uncover discrete groups of pixels in the hyperspectral PL dataset that exhibit similar spectral characteristics. Instead of analyzing gradual variations as in PCA, this approach identifies clusters of similar spectra by applying unsupervised clustering (DBSCAN) on PCA-reduced data, with t-SNE used for nonlinear dimensionality reduction and visualization.
\Cref{fig:alloy_dbscan_results}(a) shows the result of applying t-SNE to reduce the dimensionality of PCA-transformed data into two dimensions. Each point represents one pixel, and points are colored according to their assigned cluster label.
The result of \Cref{fig:alloy_dbscan_results}(a) reveals that the spectra naturally group into distinct curved branches in t-SNE space, indicating differences in similarity-based groups of PL intensities or energies. These clusters correspond to physically distinct regions in the sample, identifying $\mathrm{W}$-rich, $\mathrm{Mo}$-rich, and transition zones.
\Cref{fig:alloy_dbscan_results}(b) remarkably re-projects the same clusters identified in t-SNE space back onto the real-space coordinates of the sample. Each color corresponds to the same cluster as in \Cref{fig:alloy_dbscan_results}(a).
The spatial map in \Cref{fig:alloy_dbscan_results}(b) shows that these spectral clusters form concentric ring-like regions across the flake, consistent with a compositional gradient. This confirms that the spectral similarity strongly correlates with the physical location, reinforcing the connection between the emission properties and the alloy composition. 
This method provides a fast and objective way to visualize how different alloy compositions and emission features are distributed across the sample, pixel by pixel. Unlike manual inspection, which is slow, subjective, and prone to oversight, this automated approach immediately reveals spatial trends that would otherwise be difficult to detect.

\Cref{fig:alloy_dbscan_results}(c) shows a representative Gaussian model fit to the PL spectrum from one of the 51 selected pixels marked with black crosses in \Cref{fig:alloy_dbscan_results}(b). These spectra were fit using models with one, two, or three Gaussian components, and the fits are overlaid on the experimentally measured spectrum.
The results of \Cref{fig:alloy_dbscan_results}(c) demonstrate that a single-Gaussian model fails to fit the emission spectrum, whereas a three-Gaussian fit closely matches the measured data. This is attributed to the fact that multiple emitting quasiparticles—such as band-edge excitons and defect-bound excitons—all contribute to the observed PL signal.

\begin{figure}[tb]
    \centering
    \includegraphics[width=\linewidth]{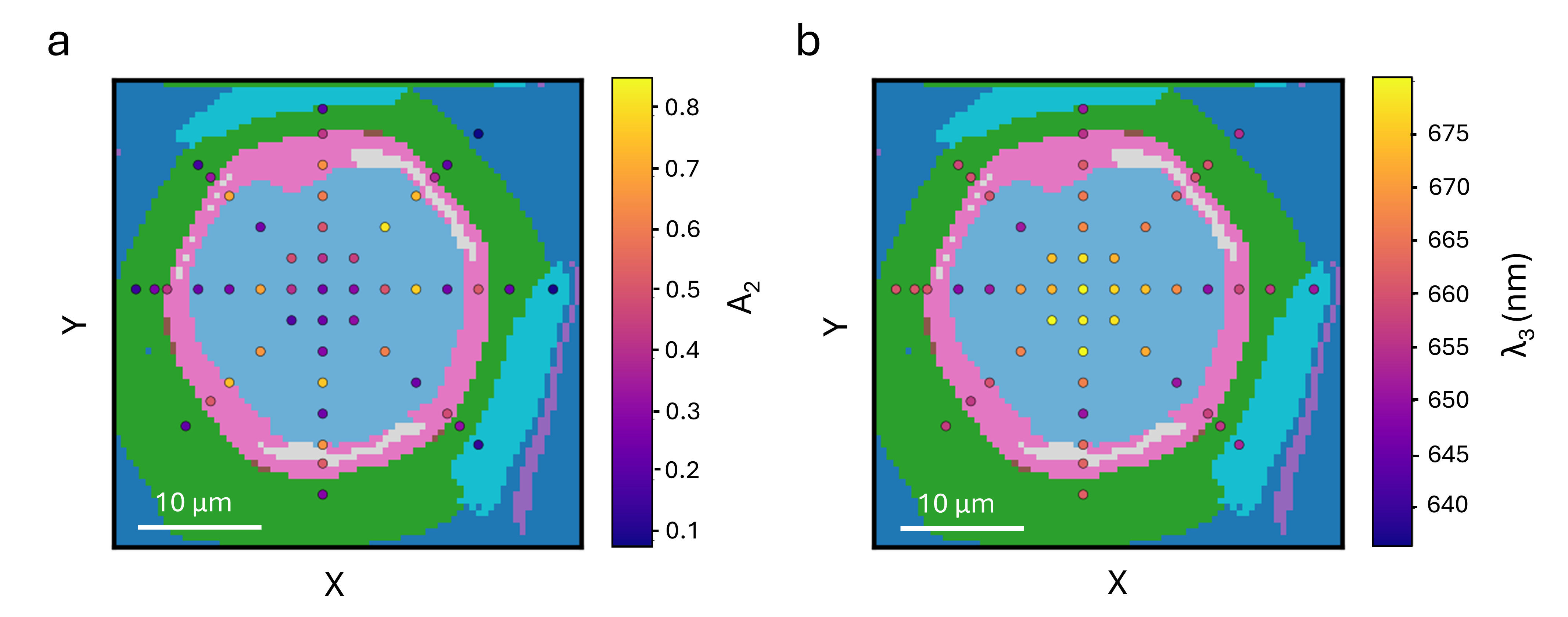}
\caption{\textbf{Spatial mapping of dominant emission parameters from three-component Gaussian fits in a graded Mo$_x$W$_{1-x}$S$_2$ monolayer.}
(a) Map of $A_{2}$ from the second Gaussian component. 
(b) Map of $\lambda_{3}$ from the third Gaussian component. 
Both maps are overlaid on the spectral cluster map from Fig.~3(b), where the 51 marked pixels (black crosses) are color-coded here according to the fitted amplitude and wavelength values.
For completeness, the full model yields six fitted parameters ($A_{1}$, $A_{2}$, $A_{3}$, and $\lambda_{1}$, $\lambda_{2}$, $\lambda_{3}$); however, only the two dominant quantities ($A_{2}$ and $\lambda_{3}$) are shown here for brevity.}
\label{fig:alloy_gaussian}
\end{figure}

\Cref{fig:alloy_gaussian} displays spatial maps of emission characteristics extracted from three-component Gaussian fits to individual PL spectra across the graded $\mathrm{Mo}_x\mathrm{W}_{1-x}\mathrm{S}_2$ flake. Each spectrum from 51 selected pixels was decomposed into three components: species~1 (band-edge exciton), species~2 (alloy disorder), and species~3 (sulfur-vacancy or defect-bound emission)~\cite{ghafariasl2024}. By amplitude, species~1 dominates in $42\%$ of pixels, reflecting strong $\mathrm{W}$-rich band-edge emission, while species~2 dominates in $50\%$, identifying alloy-disorder states as the most widespread contributors to PL intensity. Species~3 is amplitude-dominant in only $8\%$ of pixels, indicating that defect-related states rarely provide the brightest peak. By wavelength, however, species~3 dominates, appearing as the reddest feature in $86\%$ of pixels, compared to $12\%$ for species~2 and only $2\%$ for species~1. These results show that the optical response is governed primarily by species~2 in terms of intensity and by species~3 in terms of wavelength, while species~1 remains important in $\mathrm{W}$-rich domains but is generally confined to higher-energy emission.

Guided by this analysis, \Cref{fig:alloy_gaussian} maps the amplitude $A_{2}$ of species~2 and the wavelength $\lambda_{3}$ of species~3, which together capture the dominant physical signatures of the flake. In \Cref{fig:alloy_gaussian}(a), high $A_{2}$ values appear in intermediate, ring-like domains corresponding to $\mathrm{Mo}$–$\mathrm{W}$ alloy regions (such as clusters~5 and~7), while both the $\mathrm{Mo}$-rich center (cluster~8) and $\mathrm{W}$-rich rim (cluster~2) show suppressed $A_{2}$. In \Cref{fig:alloy_gaussian}(b), $\lambda_{3}$ is strongly red-shifted ($\sim$670~\text{nm}) in the central, defect-rich $\mathrm{Mo}$-dominated region (cluster~8) and gradually blue-shifts toward the outer rim (cluster~2), characteristic of cleaner $\mathrm{W}$-rich domains. Together, these spatial trends confirm that the spectral clusters identified in \Cref{fig:alloy_dbscan_results} correspond directly to physically distinct material regions and that the Gaussian-decomposed species represent unique optical transitions tied to composition and defect density across the alloy flake.
Unlike traditional PL analysis, where only a few representative spectra are manually fitted, this automated pixel-by-pixel decomposition enables comprehensive, unbiased mapping of emission species over the entire sample, capturing subtle spatial variations that would otherwise remain undetected.

\subsection{Unsupervised ML analysis of in-plane 2D $\mathrm{MoS_{2}}$–$\mathrm{WS_{2}}$ heterostructure}
Following the analysis of graded $\mathrm{Mo}_x\mathrm{W}_{1-x}\mathrm{S}_2$ alloys, where unsupervised clustering and Gaussian fitting were used to assign spectral meaning to spatial domains, we now apply a similar approach to probe the spatial and spectral heterogeneity in the 2D in-plane $\mathrm{MoS_{2}}$–$\mathrm{WS_{2}}$ sharp-interface heterostructure sample~\cite{fali2021photodegradation}.

\begin{figure}[tb]
    \centering
    \includegraphics[width=\linewidth]{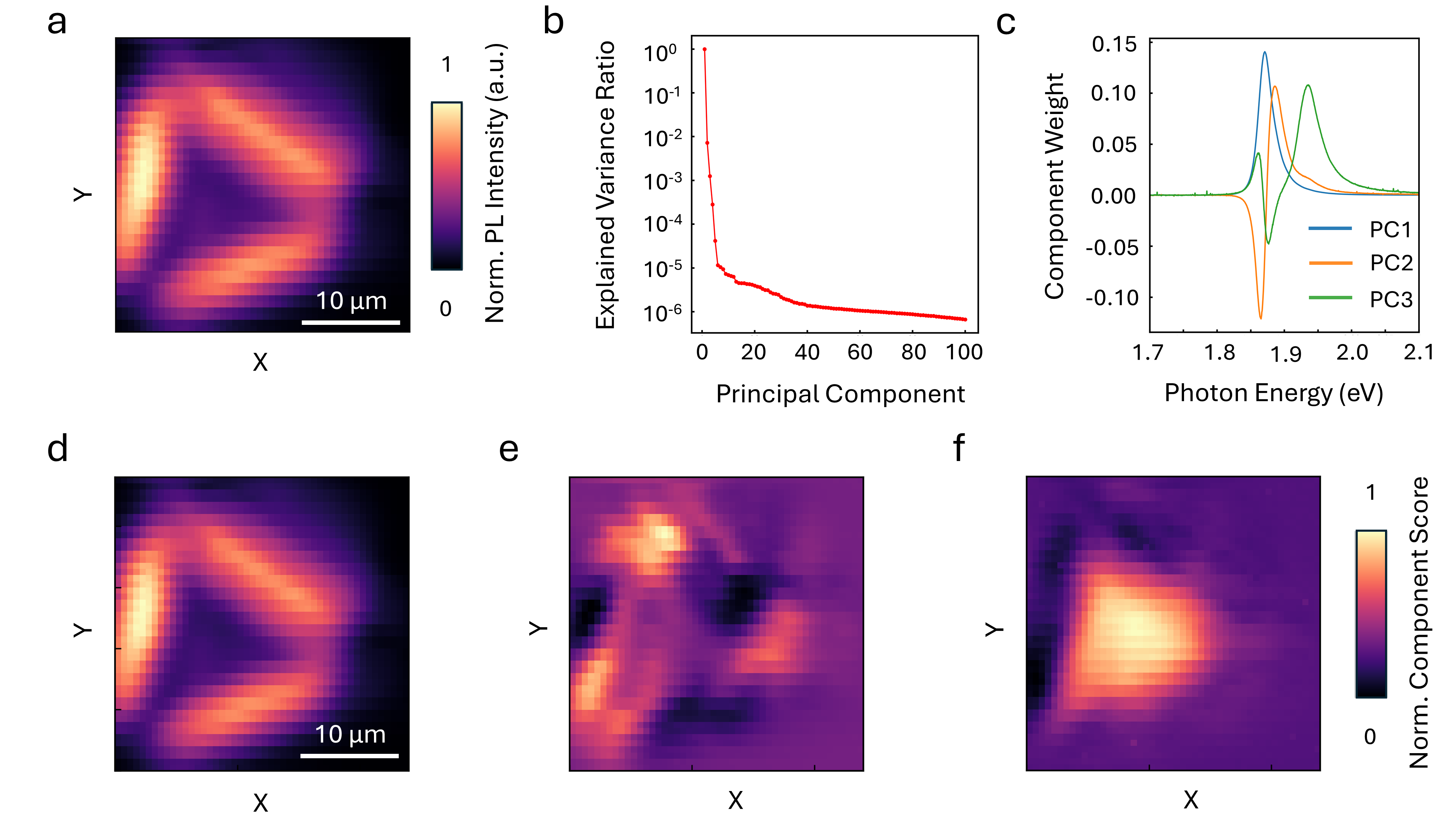}
\caption{\textbf{Hyperspectral analysis of the $\mathrm{MoS_{2}}$–$\mathrm{WS_{2}}$ heterostructure PL data.}
(a) Total PL intensity map obtained directly from experimental data (without machine learning). 
(b) PCA explained variance plot showing the fraction of total variance captured by each component. 
(c) Eigen-spectra ($\mathit{PC1}$–$\mathit{PC3}$) corresponding to the main modes of spectral variation. 
(d–f) Spatial projections of $\mathit{PC1}$–$\mathit{PC3}$ across the flake.}
\label{fig:hs_pca}
\end{figure}

\Cref{fig:hs_pca} provides a full overview of how the PL emission varies across the 2D in-plane $\mathrm{MoS_{2}}$–$\mathrm{WS_{2}}$ heterostructure. 
\Cref{fig:hs_pca}(a) Integrated PL intensity map, normalized to highlight emission contrast across the sample. \Cref{fig:hs_pca}(b) Scree plot of explained variance ratio showing that $99\%$ of the total variance is captured by the first three PCs. The corresponding PC weight vectors are plotted in \Cref{fig:hs_pca}(c). 
$\mathit{PC1}$ represents the dominant excitonic emission profile centered near $\sim$1.88–1.90~\text{eV}, reflecting the overall PL response of the system. 
$\mathit{PC2}$ shows a derivative-like shape around the same energy, which indicates spectral shifts and intensity modulations arising from local strain fields, compositional gradients, or defects. $\mathit{PC3}$ shows a distinct contribution at higher energy ($\sim$1.95~\text{eV}), pointing to additional emission associated with interfacial states or alloying effects at the heterojunction.

\Cref{fig:hs_pca}(d) reveals that the spatial distribution of $\mathit{PC1}$ resembles the integrated PL intensity map as shown in \Cref{fig:hs_pca}(a), indicating that $\mathit{PC1}$ captures the global brightness variations across the sample. 
The spatial map of $\mathit{PC2}$, as shown in \Cref{fig:hs_pca}(e), emphasizes more fragmented spatial features, particularly along boundaries and interface-like regions, suggesting $\mathit{PC2}$ represents fine-scale spectral modulations arising from strain or interfacial degradation.
\Cref{fig:hs_pca}(f) displays the spatial contribution of $\mathit{PC3}$, which is localized in a triangular central domain of the sample. This localization hints at the presence of a distinct spectral component—potentially higher-energy emission—originating from the well-defined heterointerface formed during synthesis. The triangular pattern also aligns with expected geometric boundaries of the heterojunction~\cite{fali2021photodegradation}, supporting the hypothesis that $\mathit{PC3}$ captures alloying or interfacial PL signatures unique to this structurally confined region.

\begin{figure}[tb]
    \centering
    \includegraphics[width=\linewidth]{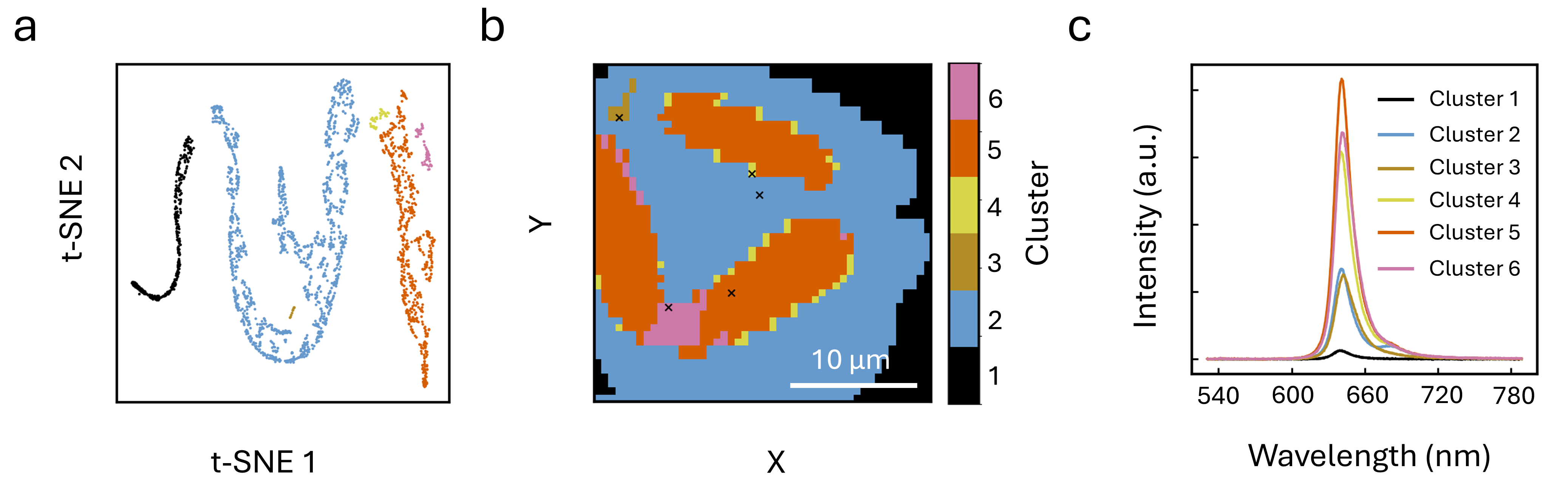}
\caption{
\textbf{Unsupervised clustering and spectral variation in the $\mathrm{MoS_{2}}$–$\mathrm{WS_{2}}$ lateral heterostructure.} 
(a) t-SNE projection of spectra colored by DBSCAN cluster labels. 
(b) Real-space mapping of the same clusters onto the sample coordinates using consistent colors; black crosses mark the pixels whose spectra are shown in panel~(c). 
(c) Representative PL spectra from the pixels marked in panel~(b), corresponding to each cluster.}
\label{fig:hs_dbscan}
\end{figure}

\Cref{fig:hs_dbscan} presents an unsupervised clustering analysis of the heterostructure hyperspectral dataset to identify spatially distinct emission domains based on spectral similarity. The result of \Cref{fig:hs_dbscan}(a) reveals that the spectra naturally organize into six well-separated branches in the t-SNE embedding space, indicating meaningful differences in PL shape, peak position, or intensity. Each cluster reflects a set of spectra that share similar emission characteristics, suggesting that the heterostructure contains multiple optically distinct regions. 
\Cref{fig:hs_dbscan}(b) reprojects the same clusters identified  onto the real-space coordinates of the sample. Each color corresponds to the same cluster label shown in the t-SNE map. 
The resulting spatial segmentation reveals that these clusters are not randomly distributed; instead, they form physically localized domains, with some clusters spanning extended interior regions while others trace narrow interfacial or boundary-like features. This spatial organization confirms that spectral similarity strongly correlates with physical location in the sample.

\Cref{fig:hs_dbscan}(c) shows representative PL spectra from individual pixels selected from each of the six clusters identified in the $\mathrm{MoS_{2}}$–$\mathrm{WS_{2}}$ heterostructure. These examples illustrate the spectral diversity across the sample. Cluster~5 exhibits a strong, broad emission centered near $640~\text{nm}$, characteristic of emission covered by pristine and alloy-stabilized $\mathrm{WS_{2}}$ domains. In contrast, Cluster~1 is nearly quenched, consistent with sulfur-deficient or degraded regions near the $\mathrm{MoS_{2}}$ side or exposed interfaces. Cluster~6 also shows strong intensity, reflecting near-pristine $\mathrm{WS_{2}}$ regions with strain broadening. Cluster~4 corresponds to interfacial areas, giving intermediate intensity with broadened line shapes. Cluster~3 appears at transitional pockets with weaker PL emission. Cluster~2 is associated with the $\mathrm{MoS_{2}}$ core but shows an asymmetric profile with a clear shoulder, indicating spectral mixing or disorder-related states. The variation among these representative spectra reflects the underlying optical heterogeneity of the heterostructure and supports the interpretation that the clusters correspond to physically distinct domains such as $\mathrm{MoS_{2}}$-rich cores, $\mathrm{WS_{2}}$-terminated edges, and strained or alloyed interface regions.

\section*{Conclusion}

The combination of high-resolution hyperspectral mapping and scalable data analysis applied to alloyed and interface-engineered two-dimensional materials provides a powerful platform to uncover fundamental physics and guide the design of next-generation optoelectronic devices.
We present an unsupervised machine learning framework that unifies PCA for dimensionality reduction, t-SNE for visualization, DBSCAN for clustering, and spectral fitting, enabling efficient decomposition of high-dimensional hyperspectral optical datasets into quantitatively interpretable features. Applying our method to complex TMDC monolayers—$\mathrm{Mo}_x\mathrm{W}_{1-x}\mathrm{S}_2$ alloys and $\mathrm{MoS_{2}}$–$\mathrm{WS_{2}}$ heterostructures—we uncovered spatially distinct optical domains, including hidden features, and linked them to compositional gradients, interfacial regions, and defect-related emission. The key strength of this approach is its ability to reveal meaningful patterns in complex datasets without requiring prior labeling or manual analysis. Beyond the specific systems studied here, the same workflow can be applied broadly to connect hyperspectral imaging data with underlying material properties, providing a versatile tool for exploring nanoscale heterogeneity in two-dimensional materials and beyond.

\begin{acknowledgement}

This work is supported by the Gordon and Betty Moore Foundation, grant DOI 10.37807/GBMF12246 and the Air Force Office of Scientific Research (AFOSR) grant number FA9550-23-1-0375.Partial support for N.T.S.
comes from the National Science Foundation (NSF) Grant
No. 2152159 (NRT-QuaNTRASE).  

\end{acknowledgement}

\begin{suppinfo}

\end{suppinfo}

\bibliography{Hyperspectrul}

@article{mak2010atomically,
  author  = {K. F. Mak and C. Lee and J. Hone and J. Shan and T. F. Heinz},
  title   = {Atomically Thin {MoS2}: A New Direct-Gap Semiconductor},
  journal = {Phys. Rev. Lett.},
  year    = {2010},
  volume  = {105},
  pages   = {136805},
  doi     = {10.1103/PhysRevLett.105.136805}
}

@article{lee2012synthesis,
  author  = {Y.-H. Lee and X.-Q. Zhang and W. Zhang and M.-T. Chang and C.-T. Lin and K.-D. Chang and Y.-C. Yu and J. T.-W. Wang and C.-S. Chang and L.-J. Li and T.-W. Lin},
  title   = {Synthesis of Large-Area {MoS2} Atomic Layers with Chemical Vapor Deposition},
  journal = {Adv. Mater.},
  year    = {2012},
  volume  = {24},
  pages   = {2320--2325},
  doi     = {10.1002/adma.201104798}
}

@article{fei2012gate,
  author  = {Z. Fei and A. S. Rodin and G. O. Andreev and W. Bao and A. S. McLeod and M. Wagner and L. M. Zhang and Z. Zhao and M. Thiemens and G. Dominguez and M. M. Fogler and A. H. Castro Neto and C. N. Lau and F. Keilmann and D. N. Basov},
  title   = {Gate-Tuning of Graphene Plasmons Revealed by Infrared Nano-Imaging},
  journal = {Nature},
  year    = {2012},
  volume  = {487},
  pages   = {82--85},
  doi     = {10.1038/nature11253}
}

@article{komsa2012two,
  author    = {Komsa, H.-P. and Krasheninnikov, A. V.},
  title     = {Two-Dimensional Transition Metal Dichalcogenide Alloys: Stability and Electronic Properties},
  journal   = {The Journal of Physical Chemistry Letters},
  year      = {2012},
  volume    = {3},
  pages     = {3652--3656},
}

@article{najmaei2013vapour,
  author  = {S. Najmaei and Z. Liu and W. Zhou and X. Zou and G. Shi and S. Lei and B. I. Yakobson and J.-C. Idrobo and P. M. Ajayan and J. Lou},
  title   = {Vapour Phase Growth and Grain Boundary Structure of Molybdenum Disulphide Atomic Layers},
  journal = {Nat. Mater.},
  year    = {2013},
  volume  = {12},
  pages   = {754--759},
  doi     = {10.1038/nmat3673}
}

@article{he2013experimental,
  author  = {K. He and C. Poole and K. F. Mak and J. Shan},
  title   = {Experimental Demonstration of Continuous Electronic Structure Tuning via Strain in Atomically Thin {Mo$S_2$}},
  journal = {Nano Lett.},
  year    = {2013},
  volume  = {13},
  pages   = {2931--2936},
  doi     = {10.1021/nl4013166}
}

@article{tongay2013defects,
  author  = {S. Tongay and J. Suh and C. Ataca and W. Fan and A. Luce and J. S. Kang and J. Liu and C. Ko and R. Raghunathanan and J. Zhou and D. F.             Ogletree and J. Li and J. C. Grossman and J. Wu},
  title   = {Defects-Activated Photoluminescence in Two-Dimensional Semiconductors: Interplay between Bound, Charged, and Free Excitons},
  journal = {Sci. Rep.},
  year    = {2013},
  volume  = {3},
  pages   = {2657},
  doi     = {10.1038/srep02657}
}

@article{huang2014lateral,
  author    = {Huang, C. and Wu, S. and Sanchez, A. M. and Peters, J. J. P. and Beanland, R. and Ross, J. S. and Rivera, P. and Yao, W. and Cobden, D. H. and Xu, X.},
  title     = {Lateral Heterojunctions within Monolayer MoSe$_2$--WSe$_2$ Semiconductors},
  journal   = {Nature Materials},
  year      = {2014},
  volume    = {13},
  pages     = {1096--1101},
}

@article{li2015epitaxial,
  author    = {Li, M.-Y. and Shi, Y. and Cheng, C.-C. and Lu, L.-S. and Lin, Y.-C. and Tang, H.-L. and Tsai, M.-L. and Chu, C.-W. and Wei, K.-H. and He, J.-H. and Chang, W.-H. and Suenaga, K. and Li, L.-J.},
  title     = {Epitaxial Growth of a Monolayer WSe$_2$--MoS$_2$ Lateral P--N Junction with an Atomically Sharp Interface},
  journal   = {Science},
  year      = {2015},
  volume    = {349},
  pages     = {524--528},
  doi       = {10.1126/science.aab4097}
}

@article{Hong2015,
  author  = {J. Hong and Z. Hu and M. Probert and K. Li and D. Lv and X. Yang and L. Gu and N. Mao and Q. Feng and L. Xie and J. Zhang and D. Wu and Z. Zhang and C. Jin and W. Ji and X. Zhang and J. Yuan and Z. Zhang},
  title   = {Exploring Atomic Defects in Molybdenum Disulfide Monolayers},
  journal = {Nat. Commun.},
  year    = {2015},
  volume  = {6},
  pages   = {6293},
  doi     = {10.1038/ncomms7293}
}

@article{Mak2016,
  author  = {K. F. Mak and J. Shan},
  title   = {Photonics and Optoelectronics of {2D} Semiconductor Transition Metal Dichalcogenides},
  journal = {Nat. Photonics},
  year    = {2016},
  volume  = {10},
  pages   = {216--226},
  doi     = {10.1038/nphoton.2015.282}
}

@article{Wang2016,
  author  = {L. Wang},
  title   = {Discovering Phase Transitions with Unsupervised Learning},
  journal = {Phys. Rev. B},
  year    = {2016},
  volume  = {94},
  pages   = {195105},
  doi     = {10.1103/PhysRevB.94.195105}
}

@article{Abate2016Nanoscopy,
  author  = {Yohannes Abate and Daniel Seidlitz and Alireza Fali and Sampath Gamage and Viktoriia Babicheva and Vladislav S. Yakovlev and Mark I. Stockman and Ramon Collazo and Dorian Alden and Nikolaus Dietz},
  title   = {Nanoscopy of Phase Separation in {In$_{x}$Ga$_{1-x}$N} Alloys},
  journal = {ACS Applied Materials \& Interfaces},
  year    = {2016},
  volume  = {8},
  number  = {35},
  pages   = {23160--23166},
  doi     = {10.1021/acsami.6b06766}
}

@article{Carrasquilla2017,
  author  = {J. Carrasquilla and R. G. Melko},
  title   = {Machine Learning Phases of Matter},
  journal = {Nat. Phys.},
  year    = {2017},
  volume  = {13},
  pages   = {431--434},
  doi     = {10.1038/nphys4035}
}

@article{Wetzel2017,
  author  = {S. J. Wetzel},
  title   = {Unsupervised Learning of Phase Transitions: From Principal Component Analysis to Variational Autoencoders},
  journal = {Phys. Rev. E},
  year    = {2017},
  volume  = {96},
  pages   = {022140},
  doi     = {10.1103/PhysRevE.96.022140}
}

@article{Ramprasad2017,
  author  = {R. Ramprasad and R. Batra and G. Pilania and A. Mannodi-Kanakkithodi and C. Kim},
  title   = {Machine Learning in Materials Informatics: Recent Applications and Prospects},
  journal = {npj Comput. Mater.},
  year    = {2017},
  volume  = {3},
  pages   = {54},
  doi     = {10.1038/s41524-017-0056-5}
}

@article{wang2018colloquium,
  author  = {G. Wang and A. Chernikov and M. M. Glazov and T. F. Heinz and X. Marie and T. Amand and B. Urbaszek},
  title   = {Colloquium: Excitons in Atomically Thin Transition Metal Dichalcogenides},
  journal = {Rev. Mod. Phys.},
  year    = {2018},
  volume  = {90},
  pages   = {021001},
  doi     = {10.1103/RevModPhys.90.021001}
}

@article{Butler2018,
  author  = {K. T. Butler and D. W. Davies and H. Cartwright and O. Isayev and A. Walsh},
  title   = {Machine Learning for Molecular and Materials Science},
  journal = {Nature},
  year    = {2018},
  volume  = {559},
  pages   = {547--555},
  doi     = {10.1038/s41586-018-0337-2}
}

@article{Choo2018,
  author  = {K. Choo and G. Carleo and N. Regnault and T. Neupert},
  title   = {Symmetries and Many-Body Excitations with Neural-Network Quantum States},
  journal = {Phys. Rev. Lett.},
  year    = {2018},
  volume  = {121},
  pages   = {167204},
  doi     = {10.1103/PhysRevLett.121.167204} 
}

@article{Djordevic2018,
  author  = {N. Đorđević and J. S. Beckwith and M. Yarema and O. Yarema and A. Rosspeintner and N. Yazdani and J. Leuthold and E. Vauthey and V. Wood},
  title   = {Machine Learning for Analysis of Time-Resolved Luminescence Data},
  journal = {ACS Photonics},
  year    = {2018},
  volume  = {5},
  pages   = {4888--4895},
  doi     = {10.1021/acsphotonics.8b01047}
}

@article{Tanaka2019,
  author  = {K. Tanaka and K. Hachiya and W. Zhang and K. Matsuda and Y. Miyauchi},
  title   = {Machine-Learning Analysis to Predict the Exciton Valley Polarization Landscape of {2D} Semiconductors},
  journal = {ACS Nano},
  year    = {2019},
  volume  = {13},
  pages   = {12687-12693},
  doi     = {10.1021/acsnano.9b04220}
}

@article{Schmidt2019,
  author  = {J. Schmidt and M. R. G. Marques and S. Botti and M. A. L. Marques},
  title   = {Recent Advances and Applications of Machine Learning in Solid-State Materials Science},
  journal = {npj Comput. Mater.},
  year    = {2019},
  volume  = {5},
  pages   = {83},
  doi     = {10.1038/s41524-019-0221-0}
}

@article{Kalantre2019,
  author  = {S. S. Kalantre and J. P. Zwolak and S. Ragole and X. Wu and N. M. Zimmerman and M. D. Stewart Jr. and J. M. Taylor},
  title   = {Machine Learning Techniques for State Recognition and Auto-Tuning in Quantum Dots},
  journal = {npj Quantum Inf.},
  year    = {2019},
  volume  = {5},
  pages   = {6},
  doi     = {10.1038/s41534-018-0118-7}
}

@article{Carleo2019_RevModPhys,
  author  = {G. Carleo and F. Becca and Y. Carrasquilla and M. Chertkov and J. Fischer and G. Kurizki and R. G. Melko and A. Turchetti and L. Tussyadiah},
  title   = {Machine Learning and the Physical Sciences},
  journal = {Rev. Mod. Phys.},
  year    = {2019},
  volume  = {91},
  pages   = {045002},
  doi     = {10.1103/RevModPhys.91.045002}
}

@article{vonLilienfeld2020,
  author  = {O. A. von Lilienfeld and K. R. Müller and A. Tkatchenko},
  title   = {Exploring Chemical Compound Space with Quantum-Based Machine Learning},
  journal = {Nat. Rev. Chem.},
  year    = {2020},
  volume  = {4},
  pages   = {347--358},
  doi     = {10.1038/s41570-020-0189-9}
}

@article{Mao2020_RamanML,
  author  = {Y. Mao and N. Dong and L. Wang and X. Chen and H. Wang and Z. Wang and I. M. Kislyakov and J. Wang},
  title   = {Machine Learning Analysis of Raman Spectra of {Mo$S_2$}},
  journal = {Nanomaterials},
  year    = {2020},
  volume  = {10},
  pages   = {2223},
  doi     = {10.3390/nano10112223}
}

@article{Frey2020,
  author  = {N. C. Frey and D. Akinwande and D. Jariwala and V. B. Shenoy},
  title   = {Machine Learning-Enabled Design of Point Defects in 2D Materials for Quantum and Neuromorphic Information Processing},
  journal = {ACS Nano},
  year    = {2020},
  volume  = {14},
  pages   = {13406--13417},
  doi     = {10.1021/acsnano.0c05267}
}

@article{fali2021photodegradation,
  author  = {A. Fali and T. Zhang and J. P. Terry and E. Kahn and K. Fujisawa and B. Kabius and S. Koirala and Y. Ghafouri and D. Zhou and W. Song and L. Yang and M. Terrones and Y. Abate},
  title   = {Photodegradation Protection in {2D} In-Plane Heterostructures Revealed by Hyperspectral Nanoimaging: The Role of Nanointerface {2D}   Alloys},
  journal = {ACS Nano},
  year    = {2021},
  volume  = {15},
  pages   = {2447--2457},
  doi     = {10.1021/acsnano.0c06148}
}

@article{Chen2021_HML_sSNOM,
  author  = {X. Chen and Z. Yao and S. Xu and A. S. McLeod and S. N. Gilbert Corder and Y. Zhao and M. Tsuneto and H. A. Bechtel and M. C. Martin and G. L. Carr and M. M. Fogler and S. G. Stanciu and D. N. Basov and M. Liu},
  title   = {Hybrid Machine Learning for Scanning Near-Field Optical Spectroscopy},
  journal = {ACS Photonics},
  year    = {2021},
  volume  = {8},
  pages   = {2987--2996},
  doi     = {10.1021/acsphotonics.1c00915}
}

@article{ElAbbassi2021_Unsupervised,
  author  = {M. El Abbassi and J. Overbeck and O. Braun and M. Calame and H. S. J. van der Zant and M. L. Perrin},
  title   = {Benchmark and Application of Unsupervised Classification Approaches for Univariate Data},
  journal = {Commun. Phys.},
  year    = {2021},
  volume  = {4},
  pages   = {50},
  doi     = {10.1038/s42005-021-00549-9}
}

@article{Kolesnichenko2021_ExcitonicSpectra,
  author  = {P. V. Kolesnichenko and Q. Zhang and C. Zheng and M. S. Fuhrer and J. A. Davis},
  title   = {Multidimensional Analysis of Excitonic Spectra of Monolayers of Tungsten Disulfide: Toward Computer-Aided Identification of Structural and Environmental Perturbations of {2D} Materials},
  journal = {Mach. Learn.: Sci. Technol.},
  year    = {2021},
  volume  = {2},
  pages   = {025021},
  doi     = {10.1088/2632-2153/abd87c}
}

@article{Joung2021_JACSAu_DLOptSpec,
  author  = {J. F. Joung and M. Han and J. Hwang and M. Jeong and D. H. Choi and S. Park},
  title   = {Deep Learning Optical Spectroscopy Based on Experimental Database: Potential Applications to Molecular Design},
  journal = {JACS Au},
  year    = {2021},
  volume  = {1},
  pages   = {427--438},
  doi     = {10.1021/jacsau.1c00035}
}

@article{Yin2021_CSR,
  author    = {Xinmao Yin and Chi Sin Tang and Yue Zheng and Jing Gao and Jing Wu and Hua Zhang and Manish Chhowalla and Wei Chen and Andrew T. S. Wee},
  title     = {Recent developments in 2D transition metal dichalcogenides: phase transition and applications of the (quasi-)metallic phases},
  journal   = {Chemical Society Reviews},
  year      = {2021},
  volume    = {50},
  pages     = {10087--10115},
  doi       = {10.1039/d1cs00236h},
}

@article{Ryu2022ML2Dmaterials,
  author  = {Ryu, Byunghoon and Wang, Luqing and Pu, Haihui and Chan, Maria K. Y. and Chen, Junhong},
  title   = {Understanding, Discovery, and Synthesis of 2D Materials Enabled by Machine Learning},
  journal = {Chemical Society Reviews},
  year    = {2022},
  volume  = {51},
  pages   = {1899--1925},
  doi     = {10.1039/D1CS00503K}
}

@article{jia2022ml_spectral_nanomat,
  author  = {H. Jia and C. Wang and C. Wang and P. Clancy},
  title   = {Machine Learning Approach to Enable Spectral Imaging Analysis for Particularly Complex Nanomaterial Systems},
  journal = {ACS Nano},
  year    = {2023},
  volume  = {17},
  pages   = {453--460},
  doi     = {10.1021/acsnano.2c08884}
}

@article{roter2022clustering,
  author  = {B. Roter and N. Ninkovic and S. V. Dordevic},
  title   = {Clustering Superconductors Using Unsupervised Machine Learning},
  journal = {Physica C},
  year    = {2022},
  volume  = {598},
  pages   = {1354078},
  doi     = {10.1016/j.physc.2022.1354078}
}

@article{Lu2022_Unraveling,
  author  = {A.-Y. Lu and L. G. P. Martins and P.-C. Shen and Z. Chen and J.-H. Park and M. Xue and J. Han and N. Mao and M.-H. Chiu and T. Palacios and V. Tung and J. Kong},
  title   = {Unraveling the Correlation between Raman and Photoluminescence in Monolayer {Mo$S_2$} through Machine-Learning Models},
  journal = {Adv. Mater.},
  year    = {2022},
  volume  = {34},
  pages   = {2202911},
  doi     = {10.1002/adma.202202911}
}

@article{Khor2023,
  author  = {J. W. P. Khor and T. T. Tran and A. S. Sharbirin and S. X. B. Yap and H. Choo and J. Kim},
  title   = {Prediction of Quantum Yields of Monolayer {$WS_2$} by Machine Learning},
  journal = {Adv. Opt. Mater.},
  year    = {2024},
  volume  = {12},
  pages   = {2302195},
  doi     = {10.1002/adom.202302195}
}

@article{ghafariasl2024,
  author  = {M. Ghafariasl and T. Zhang and Z. D. Ward and D. Zhou and D. Sanchez and V. Swaminathan and H. Terrones and M. Terrones and Y. Abate},
  title   = {Sulfur Vacancy Related Optical Transitions in Graded Alloys of {$Mo_{x}W_{1-x}S_{2}$} Monolayers},
  journal = {Adv. Opt. Mater.},
  year    = {2024},
  volume  = {12},
  pages   = {2302326},
  doi     = {10.1002/adom.202302326}
}

@article{Imamura2024,
  author  = {M. Imamura and K. Takahashi},
  title   = {Unsupervised Learning of Spatially-Resolved {ARPES} Spectra for Epitaxially Grown Graphene via Non-Negative Matrix Factorization},
  journal = {Sci. Rep.},
  year    = {2024},
  volume  = {14},
  pages   = {24200},
  doi     = {10.1038/s41598-024-73795-w}
}

@article{Fu2024_sparseSNOM,
  author  = {M. Fu and S. Xu and S. Zhang and F. L. Ruta and J. Pack and R. A. Mayer and X. Chen and S. L. Moore and D. J. Rizzo and B. S. Jessen and M. Cothrine and D. G. Mandrus and K. Watanabe and T. Taniguchi and C. R. Dean and A. N. Pasupathy and V. Bisogni and P. J. Schuck and A. J. Millis and M. Liu and D. N. Basov},
  title   = {Accelerated Nano-Optical Imaging through Sparse Sampling},
  journal = {Nano Lett.},
  year    = {2024},
  volume  = {24},
  pages   = {2149--2156},
  doi     = {10.1021/acs.nanolett.3c03733}
}

@article{lu2024when,
  author  = {B. Lu and Y. Xia and Y. Ren and M. Xie and L. Zhou and G. Vinai and S. A. Morton and A. T. S. Wee and W. G. van der Wiel and W. Zhang and others},
  title   = {When Machine Learning Meets {2D} Materials: A Review},
  journal = {Adv. Sci.},
  year    = {2024},
  volume  = {11},
  pages   = {2305277},
  doi     = {10.1002/advs.202305277}
}

@article{qi2024deep,
  author  = {Y. Qi and D. Hu and M. Zheng and Y. Jiang and Y. P. Chen},
  title   = {Deep Learning Assisted Raman Spectroscopy for Rapid Identification of {2D} Materials},
  journal = {Appl. Mater. Today},
  year    = {2024},
  volume  = {41},
  pages   = {102499},
  doi     = {10.1016/j.apmt.2024.102499}
}

@article{Zhu2024_AutonomousSTM,
  author  = {Zhu, Zhiwen and Yuan, Shaoxuan and Yang, Quan and Jiang, Hao and Zheng, Fengru and Lu, Jiayi and Sun, Qiang},
  title   = {Autonomous Scanning Tunneling Microscopy Imaging via Deep Learning},
  journal = {Journal of the American Chemical Society},
  year    = {2024},
  volume  = {146},
  number  = {42},
  pages   = {29199--29206},
  doi     = {10.1021/jacs.4c11674}
}

@article{Hillenbrand2025,
  author  = {R. Hillenbrand and Y. Abate and M. Liu and X. Chen and D. N. Basov},
  title   = {Visible-to-{THz} Near-Field Nanoscopy},
  journal = {Nat. Rev. Mater.},
  year    = {2025},
  volume  = {10},
  pages   = {285--310},
  doi     = {10.1038/s41578-024-00761-3}
}

@article{GarciadeAbajo2025Roadmap,
  author  = {F. J. García de Abajo and D. N. Basov and F. H. L. Koppens and others},
  title   = {Roadmap for Photonics with {2D} Materials},
  journal = {ACS Photonics},
  year    = {2025},
  volume  = {12},
  pages   = {3961--4095},
  doi     = {10.1021/acsphotonics.5c00353}
}

@article{He2025_OpticalID2D,
  author  = {J. He and Y. Zhou and P. Zhang and H. Zhai},
  title   = {Machine Learning Assisted Fast Optical Identification of {2D} Materials},
  journal = {Appl. Surf. Sci.},
  year    = {2025},
  volume  = {686},
  pages   = {162110},
  doi     = {10.1016/j.apsusc.2024.162110}
}
\newpage
\begin{figure}
\textbf{Table of Contents}\\
\medskip
  \includegraphics[width=\linewidth]{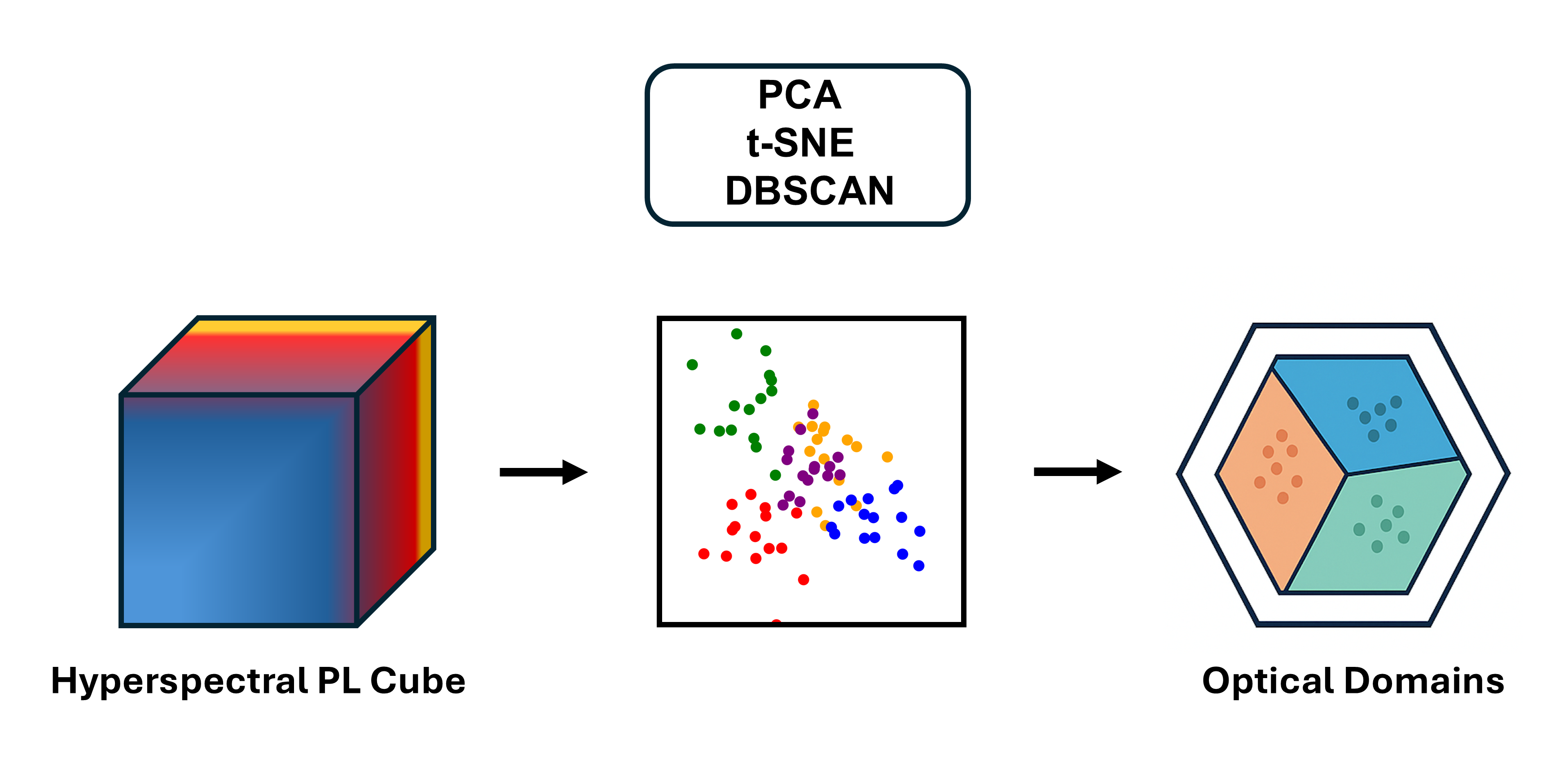}
  \medskip
  \caption*{
  Hyperspectral photoluminescence data from graded Mo$_x$W$_{1-x}$S$_2$ alloys and 
  monolayer MoS$_2$–WS$_2$ lateral heterostructures are analyzed using unsupervised 
  machine learning. The combined use of PCA, t-SNE, and DBSCAN uncovers distinct excitonic 
  regions that trace how composition, strain, and defects modulate optical responses in 
  these two-dimensional materials.
  }
\end{figure}

\end{document}